\documentclass[11pt]{article}

\usepackage{amssymb,amsfonts,amsthm,amsmath,amssymb,amscd,amstext,epsfig}
\usepackage{color}

\newcommand{\bea}{\begin{eqnarray}}
\newcommand{\eea}{\end{eqnarray}}

\def\be{\begin{equation}}
\def\ee{\end{equation}}
\def\A{A}
\def\B{B}

\def\myop{{\langle O_2 \rangle}}

\begin{document}

\begin{flushright}
PUPT-2333\\
\end{flushright}

\begin{center}
\vspace{1cm} { \LARGE {\bf An Analytic Holographic Superconductor}}

\vspace{1.1cm}

Christopher P.~Herzog

\vspace{0.7cm}

{Department of Physics, Princeton University \\
     Princeton, NJ 08544, USA }

\vspace{0.7cm}

{\tt cpherzog@princeton.edu} \\

\vspace{1.5cm}

\end{center}

\begin{abstract}
\noindent
We investigate a holographic superconductor that admits an analytic treatment near the phase transition.  In the dual 3+1 dimensional field theory, the phase transition occurs when a scalar operator of scaling dimension two gets a vacuum expectation value.  We calculate current-current correlation functions along with the speed of second sound near the critical temperature.  
We also make some remarks about critical exponents.
An analytic treatment is possible because an underlying Heun equation describing the zero mode of the phase transition has a polynomial solution.  Amusingly, the treatment here may generalize for an order parameter with any integer spin, and we propose a Lagrangian for a spin two holographic superconductor.  
\end{abstract}

\pagebreak


\setcounter{page}{1}
\setcounter{equation}{0}

\section{Introduction}

Refs.\ \cite{Gubser:2008px,Hartnoll:2008vx} observed that much of the physics of superfluids and superconductors can be captured by a simple gravitational model.  The connection, which relies on the AdS/CFT correspondence \cite{Maldacena:1997re,Gubser:1998bc,Witten:1998qj}, is that an instability for a charged black hole to develop scalar hair is dual to a superfluid or superconducting phase transition.  Because the AdS/CFT correspondence maps classical gravity to a strongly interacting field theory, the observation opens a window onto strongly interacting superconductors and superfluids where BCS theory and other weak coupling techniques are inadequate.  

The model of \cite{Gubser:2008px,Hartnoll:2008vx} has been studied over the last several years in great detail.  See \cite{Hartnoll:2009sz,Herzog:2009xv,Horowitz:2010gk} for relevant reviews.  
The model, in its most basic incarnation, involves an abelian gauge field $A_\mu$ and a charged scalar $\Psi$ in a space with a negative cosmological constant.  
Most of the studies have been numerical in nature.  Solving the equations of motion that follow from the gravity action reduces to a set of nonlinear, coupled, ordinary differential equations which can then be solved on a computer.  One notable exception is the p-wave superconductor studied
in refs.\ \cite{Basu:2008bh,Herzog:2009ci} where the dual 3+1 dimensional field theory has a vector rather than a scalar order parameter. The gravity model in this case involves a non-abelian gauge field and no charged scalar.  The order parameter is a boundary value of an off-diagonal component of the gauge field itself. 

In all of these gravity constructions, the onset of the phase transition is triggered by a zero mode for the field that gives hair to the black hole.  In the case of this 3+1 dimensional p-wave model, the zero mode can be solved for analytically, and with this solution in hand, one can study in detail properties of the field theory near the phase transition.  

It seems strange that an analytic treatment should exist for the more complicated p-wave model and not exist for the scalar model.  In this paper, we show that an analytic treatment exists for the scalar model, in particular for a 3+1 dimensional field theory and a charged scalar whose mass saturates the Breitenlohner-Freedman 
bound \cite{Breitenlohner:1982jf,Breitenlohner:1982bm}.\footnote{%
 See however refs.\ \cite{Maeda:2008ir,Gregory:2009fj} for related analytic approaches to the scalar model.
}

To a first approximation then, this paper is a rewriting of \cite{Herzog:2009ci} for the charged scalar.  We are able to compute the critical temperature $T_c$ of the phase transition.  Then near the phase transition, we calculate the speed of second sound, the change in the free energy, and the growth of the order parameter.  We can examine the phase diagram near $T_c$ as a function of superfluid velocity.  We also calculate current-current correlation functions near the phase transition and study how the hydrodynamic poles move around the complex frequency plane.
These Green's function computations offer independent confirmation of the sound speed calculated earlier in the paper.  Finally, we discuss the conductivity and London equations for the field theory.

We are also able to add a few embellishments in our rewriting of \cite{Herzog:2009ci}.  Along lines suggested by \cite{Franco:2009yz,Franco:2009if}, we consider more general potential and kinetic energy terms for the charged scalar in the action, and we study how these modifications affect the phase transition and critical exponents.  Without the modifications, the exponents are mean field.
With these modifications, we find a class of holographic superconductors with critical exponents 
$\gamma = 1$ and $1< \delta \leq 3$.  These results for the critical exponents do {\it not} appear to rely on having an analytic expression for the zero mode or 
on the dimensionality of the system.\footnote{%
While there exist earlier discussions of these critical exponents, our argument appears to be somewhat different.  
The previous literature includes refs.\ \cite{Maeda:2009wv,Iqbal:2010eh} where the authors 
make similar statements about the
exponents in the mean field case.
Refs.\  \cite{Franco:2009yz,Franco:2009if, Aprile:2009ai} discuss how the critical exponents change with modifications to the action, but largely from a numerical perspective.
}
The analytic expression for the zero mode allows us to find the constant of proportionality in the relations defining the critical exponents.  In addition to finding the critical exponents, we are able then to see in detail how the second order phase transition may be spoiled by irrelevant terms in the gravity action.  

If an analytic holographic superconductor exists for scalar and vector order parameters, then  perhaps it should exist for all integer spins.  The differential equation for the zero mode in the vector and scalar case is a linear second order differential equation with four regular singular points, i.e.\ an example of a Heun equation \cite{Ronveaux}.    The special zero mode solutions we found turn out to be Heun polynomials.  Moreover, the differential equation can indeed be generalized in a way that is suggestive of a tower of holographic superconductors with order parameters of integer spin.  We conclude by writing down a Lagrangian for a spin two holographic superconductor coupled to a gauge field.  There are well known difficulties involving 
causality, ghosts, and tachyons 
in constructing such Lagrangians, and we perform no consistency checks, leaving further investigation for future work \cite{toappear}.  
An analytic spin two holographic superconductor is clearly an object worthy of study, hopefully bringing us closer to the real world of high $T_c$ superconductors with d-wave order parameters.

\section{The Gravity Dual of a Superfluid}

In this section, we consider a holographic model of a strongly interacting, relativistic, scale invariant superfluid
\cite{Gubser:2008px, Hartnoll:2008vx,Herzog:2008he,Basu:2008st}.  
The action for a Maxwell field and a charged complex scalar field coupled to gravity is
\be
S = S_{\rm bulk} + S_{\rm bry}
\label{S}
\ee
where\footnote{%
 Roman indices $a, b, \ldots$ run from $0$ to $4$ and 
 are raised and lowered with the five dimensional metric $g_{ab}$.
 The Greek indices $\mu, \nu, \ldots$ run from $0$ to $3$ and 
 are raised and lowered with the Minkowski tensor $\eta^{\mu\nu}$ 
 with signature $(-+++)$.
}
\be
S_{\rm bulk} =  \int d^5x \, \sqrt{-g} \left[
\frac{1}{2 \kappa^2} \left( R + \frac{12}{L^2} \right) - \frac{1}{4e^2} F^{ab}F_{ab} - V- K \, | \partial \Psi - i A \Psi |^2 \right]\ .
\label{Sbulk}
\ee
and $S_{\rm bry}$ is a boundary action which ensures a well posed variational problem and also renders the on-shell action finite.  
We reveal $S_{\rm bry}$ below.  
For most of this paper, we will take 
\be
K= 1 + \A |\Psi|^2 \ , \; \; \;
V = m^2 |\Psi|^2 + \B |\Psi|^4 \ ,
\label{KV}
\ee
although in Section \ref{sec:generalsc} we consider 
slightly more general forms.\footnote{%
 We could also consider a nonminimal coupling of the form $|\Psi|^2 |F|^2$.  
 This coupling shifts the mass of the scalar field in a way that 
 spoils the analytic computation we will soon present, and so we tune it to zero by hand.
 }

This gravity action describes a dual strongly interacting 
field theory that undergoes a superfluid phase transition.
If the bulk scalar field vanishes then the equations of motion following from eq.\ \eqref{Sbulk} admit a black hole solution. Such a gravity solution corresponds to a thermal, symmetry unbroken phase of the field theory with a nonzero chemical potential. Following the general arguments in ref.\ \cite{Gubser:2008px} once the temperature of the black hole falls below a certain critical value $T_c$, a new phase exists in which the scalar field condenses. 
A solution with a nontrivial profile for $\Psi$ in the bulk 
corresponds to a superfluid phase in the field theory. 

In this paper, we work in a weak gravity (or probe) limit $\kappa / e L \ll 1$ in which gravity decouples from the scalar and gauge fields (or Abelian-Higgs sector) \cite{Hartnoll:2008vx}.  
In this limit, a solution to Einstein's equations is a black brane:
\be
ds^2 = \frac{L^2}{u^2} \left( -f(u) dt^2 + dx^2 + dy^2 + dz^2 + \frac{du^2}{f(u)} \right) \ ,
\label{metric}
\ee
where $f(u) = 1-(u/u_h)^4$. At $u=u_h>0$, there is a horizon, while in the limit $u \to 0$, the space asymptotically becomes anti-de Sitter.
We will then study the phase transition involving the Abelian-Higgs sector in this fixed background space-time.  The equations of motion in this limit are
\begin{eqnarray}
\label{scalareq}
\frac{1}{\sqrt{-g}} (\partial - i A)_a \sqrt{-g} K (\partial \Psi - i A \Psi)^a  &=& \frac{\partial V}{\partial \Psi^*} 
+ \frac{\partial K}{\partial \Psi^*}  | \partial \Psi - i A \Psi |^2 \ , \\
\frac{1}{\sqrt{-g}} \partial_c (\sqrt{-g} F^{cd} ) &=& e^2 g^{da} J_a \ ,
\label{Maxwelleq}
\end{eqnarray}
where the current is $J_a = i K [ \Psi^* (\partial - i A)_a \Psi - \Psi (\partial + i A)_a \Psi^* ]$.

From the usual rules of the AdS/CFT duality \cite{Gubser:1998bc,Witten:1998qj} the boundary value of the metric $g_{\mu\nu}$ acts as a source for the boundary theory stress tensor $T^{\mu\nu}$; the bulk gauge field $A_\mu$ acts as a source for a conserved  current $J^\mu$ corresponding to a global U(1) symmetry; and the near boundary data of the scalar $\Psi$ sources a scalar operator $O_\Delta$ with conformal scaling dimension $\Delta \geq 1$. The conformal dimension of $O_\Delta$ is related to the five dimensional mass of the scalar field $m^2$ through $m^2 L^2 = \Delta (\Delta - 4)$.  In what follows, we will be considering the particular case of $m^2 L^2 = -4$
which saturates the Breitenlohner-Freedman bound \cite{Breitenlohner:1982jf, 
Breitenlohner:1982bm}.  This mass yields an operator of dimension two.

By boundary values, we mean more specifically $\Psi^{(b)}$ and $A_\mu^{(b)}$ in the following near boundary expansions of $\Psi$ and $A_\mu$:
\begin{eqnarray}
\Psi &=&  \frac{1}{L^{3/2}} \left(  \Psi^{(b)} u^2 \ln(u/\delta) - \langle O_2 \rangle u^2 + \ldots \right)
\label{smallPsi} \ , \\
A_\mu &=& A_\mu^{(b)} + \frac{e^2}{2L} \langle J_\mu \rangle u^2 + \ldots \ , 
\label{smallAmu}
\end{eqnarray}
where we have introduced a UV cutoff $\delta \ll 1$.
%
Given these expansions, the one-point functions of the corresponding field theory operators can alternately be expressed as variations of the action with respect to the boundary values:
\be
\langle J^\mu \rangle = \lim_{u \to 0} \frac{1}{\sqrt{-g^{(b)}}} \frac{\delta S}{\delta A^{(b)}_\mu} \ , \; \; \;
\myop = \lim_{u \to 0} \frac{1}{\sqrt{-g^{(b)}}} \frac{\delta S}{\delta {\Psi^{(b)}}^*} \ ,
\label{onepoint}
\ee
where
$
g_{\mu\nu}^{(b)} = \lim_{u \to 0} (u/L)^{2} g_{\mu\nu}
$.
%
Given the fact that $\langle J^\mu \rangle$ and $A_\mu^{(b)}$ are canonically conjugate, we can identify $A_t^{(b)}$ with the chemical potential $\mu$ and $A_i^{(b)}$ with a superfluid velocity $\xi_i$.\footnote{%
 There is a minus sign discrepancy between this identification and the one in \cite{Herzog:2008he}.  Here we define $A_i(0) \equiv \xi_i$ whereas in \cite{Herzog:2008he}, $\xi_i$ was defined to be the phase gradient of the scalar.
}

Having fixed the values of $A_\mu^{(b)}$ and $\Psi^{(b)}$, we need to add certain boundary 
counter-terms $S_{\rm bry}$ to the action to have a well defined variational principle.  
There are the usual boundary cosmological constant and Gibbons-Hawking terms that are needed to regulate the Einstein-Hilbert action in an asymptotically AdS space-time.  However, we take the probe limit and our focus is on the Abelian-Higgs sector so we ignore these terms.  We do need counter-terms for the scalar:
\be
S_{\rm bry} = \left.  \int d^4{x} \, \sqrt{-\gamma} \left( 2  | \Psi |^2 / L 
+ (\Psi^* n^a \partial_a \Psi + \Psi n^a \partial_a \Psi^*) \right)
\right|_{u = \delta} \ .
\label{ctrterm}
\ee
Here $\gamma$ is the induced metric on a small radius slice $u = \delta$ and $n$ is a unit vector normal to the boundary and oriented outward.
With these counter-terms one can check that in the limit $\Psi^{(b)} \to 0$
the expansions (\ref{smallPsi}) and (\ref{smallAmu}) are consistent with the variational definitions (\ref{onepoint}) although in the computation of higher point functions further counter-terms are often required.  In the following, we find it convenient to introduce  
\be
\epsilon \equiv -\sqrt{2} \langle O_2 \rangle \ .
\ee

%

\subsection{The Superfluid Phase}

In this section, we study the superfluid phase transition for a charged scalar with
$m^2 L^2 = - 4$ in the probe limit. 

 We begin by searching for solutions to the equations of motion that are independent of 
 $x$, $y$, $z$ and $t$.  For convenience, we will set $e=1$, $L=1$ and $u_h=1$.\footnote{%
  To restore $u_h$, quantities we introduce later such as the chemical potential $\mu$, 
  frequencies $\omega$, and wave-vectors $k$ should be understood as the dimensionless   
  combinations $\mu u_h$, $\omega u_h$, and $k u_h$, respectively.
  To restore $L$ and $e$, send $\Psi \to \Psi eL $, $m \to m L$, $\A \to \A / e^2 L^2$,
  and $\B \to \B / e^2$ and 
  multiply the Abelian-Higgs action by a factor of $L/e^2$.  
  \label{footnote:units}
}
 We write the bulk scalar as $\Psi = \frac{1}{\sqrt{2}} \rho e^{i \varphi}$, and make
 a gauge transformation $A_b \to A_b + \partial_b \varphi$.  
 In the new gauge, the phase $\varphi$ disappears from the 
 equations of motion and the current becomes $J_a = K \rho^2 A_a$.  
 The $u$-component of Maxwell's equation (\ref{Maxwelleq}) now gives $K \rho^2 A_u = 0$ which means we can choose $A_u = 0$.  The equations of motion reduce to
 \begin{eqnarray}
 \label{rhoeq}
 K^{1/2} u^3 \, \partial_u \left[ \frac{f}{u^3} K^{1/2} \rho' \right] &=& \left(A_i^2 - \frac{A_t^2}{f}\right)
 \rho(1+\A \rho^2) + \frac{1}{u^2} \frac{\partial V}{\partial \rho}\ , \\
 \label{Ateq}
u \partial_u [ A_t' /u] &=& \frac{\rho^2}{u^2 f}  K  A_t \ , \\
\label{Aieq}
u \partial_u [ f A_i'/u] &=& \frac{ \rho^2}{u^2} K A_i \ ,
 \end{eqnarray}
 where $'$ indicates a derivative with respect to $u$.
 At the horizon, the boundary conditions are that $A_t(u_h) = 0$, $A_i(u_h) < \infty$, $\rho(u_h) < \infty$.  
 At $u=0$, we fix the leading values of the vector potential $A_t(0) = \mu$ and $A_i(0) = \xi_i$.
 We also set $\Psi^{(b)}=0$.
 The normal phase corresponds to the solution $A_t = \mu (1-u^2)$, $A_i = 0$, and $\rho=0$.
 
 
 To search for the superfluid phase, 
let us begin by setting $A_i = 0$.
We are interested in solutions where $\rho$ is small:
\begin{eqnarray}
\label{Atexpand}
A_t &=& \phi_0 + \epsilon^2 \, \phi_2 + \epsilon^4 \, \phi_4 + O(\epsilon^6) \ , \\
\rho &=& \epsilon \, \rho_1 + \epsilon^3 \, \rho_3 + \epsilon^5 \, \rho_5 + O(\epsilon^7) \ , 
\label{rhoexpand}
\end{eqnarray}  
where $\epsilon \ll 1$.\footnote{%
 A similar expansion was used in ref.\ \cite{Maeda:2008ir} to extract an analytic expression for the
 coherence length and to derive the London equation near the phase transition.
}
The equation for $A_t$ is solved at zeroth order by $\phi_0 = \mu_0 \, (1-u^2)$.  
In the special case $\mu_0 = 2$ and $m^2 = -4$, the equation for $\rho_1$
reduces to 
\[
\rho_1''  + \left( \frac{f'}{f} - \frac{3}{u} \right) \rho_1' +4\left( \frac{(1-u^2)^2}{f^2} + \frac{1}{u^2 f} \right) \rho_1
=0
\]
and can be solved: 
\[
\rho_1 = \frac{u^2}{1+u^2} \ .
\]
As described in Section \ref{sec:heun}, this differential equation for $\rho_1$ is related to the Heun equation, an ordinary differential equation with four regular singular points.  The existence of such a simple solution for $\rho_1$ is guaranteed by the existence of Heun polynomial solutions for certain choices of the parameters in the Heun equation.  

Our strategy in solving for this power series in $\epsilon$ will be to fix the fall-off 
$\epsilon$
but to allow the chemical potential to be corrected order by order:
$\mu = 2 + \epsilon^2 \, \delta \mu_2 + \epsilon^4 \, \delta \mu_4 + \ldots$.  Thus, in solving the differential equations, we require the boundary condition  that the $O(u^2)$ term in $\rho_i$ vanish (for $i>1$) while $\phi_i(0)$ is allowed to be nonzero.
The first few terms in the power series are 
\begin{eqnarray}
\phi_2 &=& \delta \mu_2 (1-u^2) -  \frac{u^2(1-u^2)}{8(1+u^2)} \ , \\
%
\rho_3 &=& -(1-2\A + \B) \frac{u^4}{24(1+u^2)^2}
+ (1- 8\A - 2 \B) \frac{u^2 \log(1+u^2)}{48 (1+u^2)} \ . 
%
%
\end{eqnarray}
Note that 
\be
\delta \mu_2 = \frac{1}{48} (1 + 4 \A + 4 \B) \ .
\label{deltamu2}
\ee 
If we invert the expression $\mu = \mu_c + \delta \mu_2 \, \epsilon^2 + \ldots$, we find a result for the order parameter as a function of chemical potential near the phase transition:
\be
\epsilon \approx \left(\frac{\mu - \mu_c}{\delta \mu_2}\right)^{1/2} \ .
\label{meanfieldgrowth}
\ee
If we restore dimensions, then $\mu$ should be replaced by
$\mu u_h = \mu / \pi T$.
If $\delta \mu_2 > 0$, then the new branch of solutions exists for $\mu > \mu_c$ or, fixing $\mu$ and letting $T$ vary,  
$T < T_c$.  Moreover, the order parameter grows as the square root of the reduced temperature, $\epsilon \sim (T_c - T)^{1/2}$.  This critical exponent 1/2, often called $\beta$, 
is a classic result from the Landau-Ginzburg mean field theory of phase transitions.  We will be able to confirm below that the phase transition is second order by computing the free energy difference between the phases.

The case $\delta \mu_2 < 0$ is strange.  The new branch of solutions exists only for $T>T_c$, and we will see below, in the free energy computation, that the ordered phase has a higher free energy than the normal phase.  Naively the ordered phase is unstable and the phase transition will not occur.  We suspect that $\delta \mu_2 < 0$ means one of two things.  In the first case, the free energy difference will eventually become negative if we follow this new branch of solutions far enough, and $T_c$ corresponds to a spinodal point in a first order phase transition.  In the second case, this new branch of solutions is a local maximum between the normal phase and a runaway direction in the potential landscape.  At $T_c$, the normal phase develops a runaway instability.
The terms that cause $\delta \mu_2$ to become negative are indeed problematic.  For $\B < 0$, the $\rho^4$ term in the potential is inverted.  Also, $\A < 0$ makes the kinetic term smaller.

\subsection{The free energy}
\label{sec:DeltaF}

For the free energy calculation, we also need the $\phi_4$ and $\rho_5$ terms in the expansion of $A_t$ and $\rho$.  The expressions are too cumbersome to repeat here, but we do give the near boundary expansion:
\begin{eqnarray}
\rho &=& \epsilon u^2 + O(u^4) \ , \\
A_t &=&
\left( 2 + \delta \mu_2 \epsilon^2 + \delta \mu_4 \, \epsilon^4 + O(\epsilon^6) \right) 
-
\left(2 +\left(\frac{1}{8} + \delta \mu_2 \right) \epsilon^2 + \right. \\
&& \left. \left(\frac{-5 + 37 \A + 10 \B +6(1-8 \A - 2 \B)\log 2}{1152} + \delta \mu_4 \right) \epsilon^4
%
+ O(\epsilon^6) \right) u^2
+ O(u^4) \ , \nonumber
\end{eqnarray}
where
\begin{eqnarray}
\delta \mu_4 &=& \frac{1}{276{,}480}
\bigr( -1265 + 16{,}568 \A+ 3320 \B  - 62{,}960 \A^2 - 31{,}600 \A \B-4820 \B^2 
  \nonumber \\
&&
\hspace{20mm}+ 240(1 - 8 \A - 2 \B)(7 - 44 \A - 8 \B) \log 2 \bigl) \ . 
\end{eqnarray}

The free energy of the field theory 
is determined by the value of the
action (\ref{S}) evaluated on-shell, 
$\Omega = -T S_{\rm os}$. 
 Employing the equations of 
motion, we may rewrite the bulk action (\ref{Sbulk}) (in the case $\A = \B = 0$) as
\be
S_{\rm bulk} =  
\int d^{4}x \left[ \frac{\sqrt{-g} g^{uu}}{2} \left. 
 \left(g^{\mu\nu} A_\mu A_\nu'
+ \rho \rho' \right) \right|_{u=\delta} +\frac{1}{2} 
 \int_\delta^{u_h} du \sqrt{-g} A_\mu A^\mu  \rho^2 \right] \ .
\ee
We have included a cut-off $\delta$ to regulate the boundary divergences.  
For general $\A$ and $\B$, the integral over $u$ has more terms which we do not write.
Using the explicit form of the metric (\ref{metric}) and the boundary counter-terms introduced
above (\ref{ctrterm}), 
the on-shell action reduces to
\be
S_{\rm os} =  \frac{1}{2T} \, \mbox{Vol}_3 \left[ \mu J^t + \xi_i J^i   +
 \int_0^{u_h}  du\,  \sqrt{-g} \,  A_\mu A^\mu
 \rho^2  \right] \ .
\ee
With our homogenous ansatz for the fields, the integral over $d^4x$ yields a factor of 
$\mbox{Vol}_3 / T$ where $\mbox{Vol}_3$ is the spatial volume of the field theory.

For a background where $\Psi = 0$ and
\be
A_t = (2 + \delta \mu_2 \, \epsilon^2 + \delta \mu_4 \, \epsilon^4 + O(\epsilon^6)) (1-u^2) \ ,
\ee
the on-shell action is
\begin{eqnarray}
S_{\rm vac} 
&=& \beta \, \mbox{Vol}_3 \Bigl(4+ \frac{1}{12} (1+4 \A + 4 \B)\epsilon^2  \\
&&
+ 
\frac{1}{69{,}120}
\bigl(
-1235 + 16{,}808 \A+ 3560 \B - 62{,}480 \A^2  - 30{,}640 \A \B - 4340 \B^2 
\nonumber
\\
&&
\hspace{20mm} + 240(1-8 \A - 2 \B)(7 - 44 \A - 8 \B) \log 2 \bigr) \epsilon^4 +
O(\epsilon^6) \Bigr) \ . \nonumber
\end{eqnarray}
 For the background where the scalar has condensed and hence $\Psi \neq 0$, we find in contrast that
\begin{eqnarray}
S_{\rm sf} &=& \beta \, \mbox{Vol}_3 \Bigl(4+ \frac{1}{12} (1+4 \A + 4 \B)\epsilon^2  \\
&&
+ 
\frac{1}{69{,}120}
\bigl(
-1055 + 17{,}258 \A + 4280 \B  - 62{,}480 \A^2 - 30{,}640 \A \B- 4340 \B^2 
\nonumber
\\
&&
\hspace{20mm} + 240(1-8 \A - 2 \B)(7 - 44 \A - 8 \B) \log 2 \bigr) \epsilon^4 +
O(\epsilon^6) \Bigr) \ . \nonumber
\end{eqnarray}
Now
\be
\Delta \Omega = \Omega_{\rm sf} - \Omega_{\rm vac} = T (S_{\rm vac} - S_{\rm sf} ) 
= {\rm Vol}_3 \left( - \frac{1}{384}(1+4 \A + 4 \B) \epsilon^4 + O(\epsilon^6) \right) \ .
\label{deltaomega}
\ee

Intriguingly, the sign of $\Delta \Omega$ is correlated with the sign of $\delta \mu_2$ (\ref{deltamu2}).  Fixing $\mu$ and letting $T$ vary, 
when the ordered phase exists for $T>T_c$, the free energy difference is positive, and the ordered phase is less stable.
When the ordered phase exists for $T<T_c$, the free energy difference is negative, and the ordered phase is more stable.  
From the fact that the free energy difference scales as $\epsilon^4$, we see that the phase transition is second order.  For small $\epsilon$, 
$\epsilon^4 \sim (\mu - \mu_c)^2$. Thus, $\epsilon^4 \sim (T_c - T)^2$.  The derivative of $\Delta \Omega$ with respect to temperature is continuous but non-differentiable. 
There exists a critical exponent $\alpha$ which describes the behavior of the specific heat
$C_v = T \, \partial S / \partial T \sim (T_c - T)^{-\alpha}$ near a phase transition.  From the behavior of the free energy, we see that $\alpha = 0$, consistent with Landau-Ginzburg mean field theory.

\subsection{Superfluid Flow}
\label{sec:superfluid}

In this section, we generalize the background above to allow for the possibility of a superfluid flow.  In terms of the bulk solution, this generalization requires turning on a constant value of $A_i(0)= \xi_i$.  Invoking rotational invariance, we take $\xi_i = (\xi,0,0)$ without loss of generality.  We again solve the system in a small $\epsilon$ expansion but we treat $\xi$ as a small parameter.  In this double series expansion, we find that
\begin{eqnarray}
A_t &=& \left( 2 + \frac{1}{2} \xi^2 \right)(1-u^2) +
\phi_2 \,  \epsilon^2 + \ldots \ ,
 \\
A_x &=& \xi \left( 1 - \frac{u^2 }{8(1+u^2)} \epsilon^2 \right) + \ldots \ , \\
\rho &=&  \epsilon \left(\frac{ u^2}{1+u^2} - \xi^2 \frac{ u^2 \ln (1+u^2)}{4 (1+u^2)} \right) + \ldots \ .
\end{eqnarray}
These solutions can be used to compute the speed of second sound.  In a two component fluid, there are typically two propagating collective modes, ordinary and second sound.  In our probe approximation the two sounds decouple \cite{Herzog:2008he}; second sound is associated purely with the Abelian-Higgs sector while ordinary sound is associated only with the gravitational sector.

The speed of second sound, like that of ordinary sound, can be computed from derivatives of the state variables.    From ref.\ \cite{Herzog:2008he}, the second sound speed squared in this probe limit should be
\be
c_2^2 = \left. -\frac{ \partial J^x / \partial \xi}{\partial J^t / \partial \mu} \right|_{\xi=0} \ .
\label{speedsecondsound}
\ee
The small $u$ expansion for the vector field is
\begin{eqnarray}
A_t &=& \mu + \frac{12-\mu(7 + 4 \A + 4 \B)}{1+4 \A + 4 \B} u^2 + \ldots\ , 
\label{Atsfexpand}
\\
A_x &=& \xi - \frac{6\xi (\mu - 2)}{1+ 4\A + 4 \B} u^2 + \ldots \ .
\label{Axsfexpand}
\end{eqnarray}
Hence, up to higher order corrections in $\epsilon$,
\be
c_2^2 \approx \frac{6}{7+4 \A + 4 \B} (\mu - 2) \approx \frac{ \epsilon^2}{8} \frac{ 1+ 4 \A + 4 \B}{ 7 + 4 \A + 4 \B} \ .
\label{secsoundthermo}
\ee

This perturbative solution in $\xi$ can also be used to analyze the phase diagram of the system near the critical point $\mu_c = 2$.  At the critical point, we expect the order parameter to vanish, so $\epsilon =0$.  The value of $A_t$ at $u=0$ can be reinterpreted as the value of the chemical potential.  These two facts give us a relation between the chemical potential and superfluid velocity along the critical line separating the two phases.  We expect that
\be
\mu \approx 2 + \frac{1}{2} \xi^2 \ .
\ee

\subsection{More General Superconductors and Critical Exponents}
\label{sec:generalsc}

In this section, following a suggestion of \cite{Franco:2009yz, Franco:2009if}, we may consider a more general form for $V$ and $K$.  Consider
\be
K(\rho) = 1  + \A \rho^{a-2}\ ; \; \; \;
V(\rho) = \frac{1}{2} m^2 \rho^2 + \B \rho^a \ ,
\label{KValt}
\ee
where $2<a<4$.
Then we need to consider a more general expansion for the fields of the form
\begin{eqnarray}
\label{Atexpandgen}
A_t &=& \mu_c (1-u^2) + \epsilon^{a-2} \, \phi_{a-2}  + \ldots \ , \\
\label{rhoexpandgen}
\psi &=& \epsilon \, \rho_1 + \epsilon^{a-1} \, \rho_{a-1}  + \ldots\ . 
\end{eqnarray}
We find that $\phi_{a-2} = \delta \mu_{a-2} (1-u^2)$, and hence
\be
\epsilon = \left( \frac{\mu-\mu_c}{\delta \mu_{a-2}} \right)^{1/(a-2)} \ .
\ee
Thus, we can get any critical exponent $\beta$ we desire between $1/2$ and $\infty$ by tuning the value of $a$ and making sure that $\delta \mu_{a-2} > 0$.  As in the case $\delta \mu_2 < 0$, for $\delta \mu_{a-2} < 0$, we anticipate a first order phase transition or runaway pathologies.  If we want a critical exponent less than 1/2, we first have to tune the values of quartic contributions such that $\delta \mu_2 = 0$.  Then we have the possibility of adding higher order terms with $a>4$.  

 In the particular case $a=3$, we find $\delta \mu_{1} =  (8 \A + 9 \B)/8$ and
\be
\rho_{2} =
-(4 \A + 3 \B) \frac{u^2 \log(1+u^2)}{8 (1+u^2)} 
\ .
\ee
We have also calculated the free energy difference for this model, using the method detailed in section \ref{sec:DeltaF}.  The result is 
\be
\Delta \Omega = \Omega_{\rm sf} - \Omega_{\rm vac} 
= {\rm Vol}_3 \left( - \frac{1}{192}
\left(8 \A (5 - 6\log 2) + 9 \B ( 3 - 4 \log 2) \right) \epsilon^3 + O(\epsilon^4) \right) \ .
\label{deltaomega2}
\ee
We see that $\Delta \Omega \sim (T_c - T)^{3}$ and thus the critical exponent $\alpha = -1$.

So far, we have focused on two critical exponents, $\alpha$ and $\beta$, describing the behavior of the specific heat and the order parameter near the phase transition.  There are several more, although general scaling arguments demonstrate that only two of these exponents are independent.  The rest are fixed through scaling relations such as 
$\alpha = 2 - \beta(\delta + 1)$ and $\gamma = \beta (\delta -1)$.  The exponent $\delta$ is defined at $T=T_c$ through the relation
\be
\Psi^{(b)} \sim  \langle O_2 \rangle^\delta
\label{deltascale}
\ee
while $\gamma$ is the susceptibility for $T \lesssim T_c$:
\be
\frac{\partial \langle O_2 \rangle}{\partial \Psi^{(b)}} \sim (T_c - T)^\gamma \ .
\ee

As we have seen above, $\beta$ is fixed by the form of the series expansion (\ref{Atexpandgen}) and (\ref{rhoexpandgen}), and in particular by the subleading term in $A_t$.  The critical exponent $\delta$ is fixed in a similar fashion, by the subleading term in $\rho$.  In the previous examples, we solved for $\rho_{a-1}$ by insisting on three conditions: that $\rho_{a-1}$ be finite at the black hole horizon and that the leading and subleading terms at the boundary vanish.  In general, we would not be able to impose three conditions on a second order linear differential equation, but we were able to tune the value of $\delta \mu_{a-2}$ that we had left arbitrary in solving for $\phi_{a-2}$.  
We could have solved this system in a slightly different way, setting $\delta \mu_{a-2}=0$, and thus freezing $\mu / T$ to the critical value.  Still imposing regularity at the horizon, we would find that we need to turn on a source $\Psi^{(b)}$ for the operator $O_2$ in the solution for $\rho_{a-1}$.  Hence, there would be
a scaling relation of the form
 (\ref{deltascale}) where $\delta = a-1$.  
 
For the class of models we are considering, we see that $\beta = 1/(a-2)$ and $\delta = a-1$.  From the scaling relations, the susceptibility will always be $\gamma = 1$ while the heat capacity is described by
\be
\alpha = \frac{a-4}{a-2} \ .
\ee
This value for $\alpha$ agrees with our two free energy calculations in the cases $a=3$ and $a=4$.
It would be interesting to verify this relation explicitly for all $a$ such that $2 < a \leq 4$, but we have no reason to doubt its veracity.  These results 
appear to be independent of the space-time dimension and 
of whether we have an explicit analytic solution for $\rho_1$.  What knowledge of $\rho_1$ provides are the constants of proportionality in the definitions of the critical exponents and 
confidence that the series expansion in $\epsilon$ is well behaved.

Previous papers discussing critical exponents in a similar context are
\cite{Franco:2009yz,Franco:2009if,Maeda:2009wv,Iqbal:2010eh,Aprile:2009ai}.


\section{Green's Functions}
\label{sec:greens}

To calculate the Fourier transformed retarded current-current correlation functions, we need to study fluctuations of the gauge and scalar field in our background.  Our background is the black D3-brane (\ref{metric}) with chemical potential $\mu \gtrsim \mu_c$ close to the critical value and no superfluid velocity.  (There is no obstruction to introducing a small nonzero $\xi \neq 0$ but the analysis becomes more complicated.)

We will consider two types of fluctuations, 
a transverse fluctuation $\delta A_x \sim e^{-i \omega t + i k y}$ and a longitudinal fluctuation $\delta A_x \sim e^{-i \omega t + i k x}$.  
The first type of fluctuation contains no propagating modes while the second will allow us to look at the second sound and shear modes.  To keep the analysis simple, we restrict to the hydrodynamic regime where the order parameter, frequency, and wave-vector are small compared to the temperature.  In our dimensionless notation, $\epsilon$, $k$, $\omega \ll 1$.  
We will also set $\A = \B = 0$ in the following, choosing the simplest possible forms for 
$V$ and $K$ (\ref{KV}). 

\subsection{Pure Transverse Mode}
\label{sec:transverse}

The transverse mode is described by fluctuations of the field $A_x$ with only $y$ spatial dependence.  We decompose the fluctuations into Fourier modes:
\be
\delta A_x(u,t,y) = a_x(u) e^{-i \omega t + i k y} \ .
\ee
These fluctuations decouple from the other scalar and gauge field components and satisfy the second order differential equation, 
\be
a_x'' + \left(\frac{f'}{f} - \frac{1}{u} \right)a_x' + \left( \frac{ \omega^2 - k^2 f}{f^2} - \frac{\rho^2}{u^2 f} \right) a_x = 0 \ .
\label{sheardiffeq}
\ee

Near the horizon $u=1$, we find that $a_x \sim (1-u)^{\pm i \omega/4}$ satisfies either ingoing or outgoing plane wave type boundary conditions.  Consistent with the presence of an event horizon, it is natural to choose ingoing boundary conditions (the minus sign in the exponent).  This choice leads to retarded, as opposed to advanced, Green's functions 
in the dual field theory \cite{Son:2002sd}.  At the boundary $u=0$ of $AdS$, we would like the freedom to set $a_x(0) = a_{x0}$ to some arbitrary value of our choosing, corresponding to perturbing the dual field theory by a small external field strength.  These two boundary conditions along with the differential equation uniquely specify the functional form of $a_x$.  

While an analytic solution of (\ref{sheardiffeq}) does not appear to be available, one can easily solve this equation in the limit of small $\omega$, $k$, and $\epsilon$.  The first few terms in this expansion are
\be
a_x = a_{x0} \left( \frac{1-u^2}{1+u^2} \right)^{-i \omega/4} \left(1 + \epsilon^2 a_{x \epsilon}
+ \epsilon^2 \omega a_{x \omega \epsilon} + k^2 a_{xk} + \omega^2 a_{x \omega} + \ldots \right) \ .
\label{axexpand}
\ee
We find
\begin{eqnarray}
a_{x\epsilon} &=& - \frac{ u^2 }{8(1+u^2)} \ , \\
a_{x \omega \epsilon} &=& - \frac{i u^2 }{16 (1+u^2)} \ , \\
a_{x k} &=& \frac{1}{4} \ln (u) \ln \left( \frac{1+u^2}{1-u^2} \right) - \frac{1}{4} \mbox{Li}_2(u^2) + \frac{1}{16} \mbox{Li}_2(u^4) \ .
\end{eqnarray}
The expression for $a_{x \omega}$ is too cumbersome to give here.  Near the boundary, this solution (\ref{axexpand}) has the expansion,
\be
a_x = a_{x0} + a_{x0} \left(
\frac{i \omega}{2} 
-\frac{ \epsilon^2}{8} - \frac{i\omega \epsilon^2}{16}
- \frac{\omega^2 \log 2}{4} 
+\frac{1}{2}(\omega^2 - k^2) \left( \frac{1}{2} - \ln (u) \right)
\right) u^2 + \ldots \ .
\ee
From this near boundary expansion, we can calculate the two-point function for the current in the hydrodynamic limit:
\be
G^{xx}(\omega, k) = 2 \left( 
\frac{i \omega}{2} 
-\frac{ \epsilon^2}{8} - \frac{i\omega \epsilon^2}{16}
- \frac{\omega^2 \log 2}{4} 
+(\omega^2 - k^2) c
\right) \ .
\ee
This Green's function is morally 
a second derivative of the on-shell action with respect to $a_{x0}$.
Note that there is a counter-term ambiguity, proportional to an arbitrary constant $c$, of the form discussed in \cite{Herzog:2009ci,Skenderisreview}.

\subsection{Second Sound Mode}
\label{sec:secondsound}

In general, second sound modes are expected to produce poles in the density-density correlation function.  We thus need to consider fluctuations in the conjugate field $A_t$.  
A self-consistent set of fluctuation equations also involves the scalar field and longitudinal fluctuations of $A_x$:
\begin{eqnarray}
\delta A_t(u,t,x) &=& a_t(u) e^{-i \omega t + i k x} \ , \\
\delta A_x(u,t,x) &=& a_x(u) e^{-i \omega t + i k x} \ , \\
\delta \Psi(u,t,x) &=& \psi(u) e^{-i \omega t + i k x} / \sqrt{2} \ .
\end{eqnarray}
These four fluctuating modes satisfy  four second order differential equations 
(if one also counts the equation for $\psi^*$) and one first order constraint equation:
\begin{eqnarray}
a_t'' - \frac{1}{u} a_t' - 
\left( k^2  + \frac{ \rho^2}{u^2} 
\right) \frac{a_t}{f} - \frac{ k \omega}{f} a_x -
\frac{ \rho}{2 u^2 f}\left( (\omega + 2 \phi) \psi - (\omega - 2 \phi) \psi^* \right) &=& 0 \ , 
\label{ateq} 
\\
a_x'' + \left( \frac{f'}{f} - \frac{1}{u} \right) a_x' 
+ \left( \frac{ \omega^2}{f^2}  - \frac{ \rho^2}{u^2 f} \right) a_x 
 + \frac{ k \omega}{f^2} a_t +
 \frac{  k \rho (\psi - \psi^*)}{2 u^2 f} &=& 0\ , \\
\psi'' + \left( \frac{f'}{f} - \frac{3}{u} \right) \psi' + \left( \frac{(\omega + \phi)^2}{f^2} + \frac{4 - k^2 u^2}{u^2 f} \right) \psi + \frac{ (\omega + 2 \phi) \rho}{ f^2} a_t  + \frac{k \rho}{f}a_x &=& 0 \ , \\
\omega a_t' + k f a_x' + \frac{ f}{2 z^2}\left(
(\psi^*-\psi) \rho' - ({\psi^*}' - \psi') \rho 
\right) &=& 0 \ .
\label{constraint}
\end{eqnarray}
We have checked that the derivative of the constraint equation (\ref{constraint}) with respect to $u$ is a linear combination of all five differential equations (\ref{ateq})--(\ref{constraint}).  Thus if a solution of the first four differential equations satisfies the constraint for some $u$, it will satisfy the constraint for all $u$. 

There are seven integration constants associated with this linear system (\ref{ateq})--(\ref{constraint}).  If we look at the horizon of the black hole at $u=1$, we find seven different kinds of behavior.  There exist six solutions that have plane wave behavior for $a_x$, $\psi$, and $\psi^*$ near the horizon of the form $(1-u)^{\pm i \omega/4}$.  There is also a pure gauge solution,
\be
a_t = -i \omega \ , \; \; \;
a_x = i k \ , \; \; \;
\psi = i \rho \ , \; \; \;
\psi^* = - i \rho \ .
\ee
As in the analysis for the transverse fluctuations, we choose pure ingoing boundary conditions corresponding to $(1-u)^{-i\omega/4}$ behavior.  At the boundary $u=0$ of our asymptotically AdS space, we would like to be able to perturb the system with arbitrary boundary values of $a_t$ and $a_x$ but set the source terms for $\psi$ and $\psi^*$ to zero.  These are four constraints and we have only three ingoing solutions.  Thus we will also need to make use of the pure gauge solution to enforce the boundary conditions at $u=0$. 

We solved this system perturbatively in $\omega$, $k$, and $\epsilon$.  We present the results here in the limit where $\omega \sim k^2 \sim \epsilon^2$.  The near boundary expansion $(u=0)$ of the solution takes the form
\begin{eqnarray}
a_t &=& a_{t0} + k \frac{a_{t0} k + a_{x0}\omega}{{\mathcal P}}
\left[48 i k^4 + 7 i  \epsilon^4 + 148  \epsilon^2 \omega - 480 i \omega^2 
\right. \nonumber \\
&& 
\left. \hspace{50mm }+ 8 k^2 (5 i  \epsilon^2 + 12 \omega)
\right] u^2 + \ldots
\label{atexpand}
\ , \\
a_x &=& a_{x0} - \omega \frac{a_{t0} k + a_{x0}\omega}{{\mathcal P}}
\left[ 
48 i k^4 + 7 i  \epsilon^4 + 148  \epsilon^2 \omega - 480 i \omega^2 
\right. \nonumber \\
&& 
\left. \hspace{50mm }+ 8 k^2 (5 i  \epsilon^2 + 12 \omega)
\right] u^2 + \ldots
\ , \\
\psi &=& -\frac{a_{t0}}{{\mathcal P}} 8 i \epsilon 
\left[ 12 k^4 + (7  \epsilon^2 - 120 i \omega) \omega + 3 k^2 (  \epsilon^2 + 16 (1-i) \omega) \right] u^2 
\nonumber \\
&&
-\frac{a_{x0}}{{\mathcal P}} i \epsilon k
\left[ 48 k^4 +  \epsilon^4 +12(2-3i) \epsilon^2 \omega - 
96 ( 3 + 4 i) \omega^2 \right.
\nonumber \\
&&  \hspace{25mm} \left. + 16 k^2 ( \epsilon^2 + 6(1-2i)\omega)) \right] u^2 + \ldots \ ,
\end{eqnarray}
where the pole structure is given by 
\begin{eqnarray}
{\mathcal P} &=& 
960 \omega^3 + 56i (12 k^2 +  \epsilon^2) \omega^2 
-12 (16 k^2 + 3  \epsilon^2) k^2 \omega \nonumber \\
&&
- i (48 k^4 + 16 k^2  \epsilon^2 +  \epsilon^4 ) k^2 \ .
\end{eqnarray}
Note the expression $a_{t0} k + a_{x0} \omega$ is not homogenous in our scaling limit.  We leave this piece in the Green's functions to make it obvious that the Green's functions satisfy the Ward identity $k_\mu G^{\mu\nu} = 0$.  We have included the leading corrections proportional to $a_{t0}$ and $a_{x0}$.  There are terms in the expansion proportional to $u^2 \ln u$ that give rise to the counter-term subtraction ambiguity mentioned above but they are subleading in $\omega$, $k$, and $\epsilon$.

Let us study the pole structure in two different limits.  In the limit $k\ll \epsilon$, the poles are located at 
\begin{eqnarray}
\omega &=& \pm \frac{ k  \epsilon}{2 \sqrt{14}} - \frac{33 i k^2}{196} \ , \\
\omega &=& - \frac{ 7 i  \epsilon^2}{120} - \frac{89 i k^2}{245} \ .
\end{eqnarray}
The first two poles are propagating modes that we identify with second sound.  Indeed, the speed of second sound agrees with the earlier result (\ref{secsoundthermo}) from Section \ref{sec:superfluid}.  The position of the third pole is determined mostly by the size of the order parameter $\epsilon$ and so we associate it with the zero mode that causes the phase transition from the superfluid phase back to the normal phase.

In the opposite limit, $k \gg \epsilon$, where the order parameter is small, the behavior should be close to that of the normal fluid.  In this limit, we find
\begin{eqnarray}
\omega &=& \frac{\pm 3 - i}{10} k^2 + \frac{\pm 9 - i }{ 240}  \epsilon^2 \ , \\
\omega &=& - \frac{ i k^2}{2} - \frac{ i  \epsilon^2}{20} \ .
\end{eqnarray}
The first two poles are associated with the zero modes that cause the phase transition from the normal phase to the superfluid phase, while the third pole is associated to a diffusive mode of the conserved charge density.  Indeed the location of this diffusive pole is determined by the dynamics of the normal phase and was calculated without the order $\epsilon^2$ correction, long ago
\cite{Policastro:2002se}.  As we vary $\epsilon$ and $k$, the number of poles cannot change.  The two zero mode poles evolve into the sound poles of the previous limit while the diffusive pole becomes the zero mode pole of the previous limit.

As a further check, we consider a particular static limit of the density-density correlation function.  From (\ref{atexpand}), we can read off the Green's function,
\be
G_{tt}(\omega, k) = - 2 \frac{k^2}{\mathcal{P}} \left(48 i k^4 + 7 i  \epsilon^4 + 148  \epsilon^2 \omega - 480 i \omega^2+ 8 k^2 (5 i  \epsilon^2 + 12 \omega) \right)  \ .
\ee
We are interested in the long wave-length limit of this Green's function:
\be
\lim_{k \to 0} G^{tt}(0,k) = 14 \ .
\label{longGtt}
\ee
This long wave-length limit is equal to a thermodynamic susceptibility,
\be
\lim_{k \to 0} G^{tt}(0,k) = \frac{\partial^2 P}{\partial \mu^2} = \frac{\partial J^t}{\partial \mu} \ .
\ee
Given this relation, we see that (\ref{Atsfexpand}) agrees with (\ref{longGtt}). 

\subsection{Ohm's Law and London Equations}

Via Ohm's Law, a homogenous, time dependent electric field $\delta A_x \sim e^{-i \omega t}$ should produce a current in the system.  To investigate this long wave-length limit of the conductivity, we set $k=0$ for the two-point functions computed above.  We have
\be
 G^{xx}(\omega, 0) =2 \left( - \frac{ \epsilon^2}{8} + i\left( \frac{1}{2} - \frac{ \epsilon^2}{16} \right) \omega + c \, \omega^2  + O(\omega^3) \right) \ . 
\label{Gxx}
\ee
Note that the $k \to 0$ limits of $G^{xx}$ computed in Sections \ref{sec:transverse} and \ref{sec:secondsound} agree. 

Identifying the electric field $E_x = i \omega \delta A_x$ and recalling Ohm's Law, the conductivity is related via a Kubo type formula to the retarded Green's function
\be
\sigma_{xx} (\omega) = \frac{G^{xx}(\omega)}{i \omega} \ .
\ee
The term proportional to $\epsilon^2$ in $G^{xx}$ thus produces a pole in the imaginary part of the conductivity.  As discussed in \cite{Hartnoll:2008vx, Hartnoll:2008kx}, by the Kramers-Kronig relations (or by properly regularizing the pole), there must be a delta function in the real part of the conductivity, indicating the material loses all resistance to DC currents and suggesting the phase transition is to a superconducting state.  Here we can calculate the residue of the pole analytically close to the phase transition:
\be
\mbox{Res}_{\omega=0} \sigma_{xx} = 2 \frac{i  \epsilon^2}{8} \ .
\ee

An important observation is that in our system, the $\omega \to 0$ and $k \to 0$ limits of the Green's functions commute.  The residue of the pole in the conductivity is related to the long wavelength limit of the current-current correlation function in the following way:
\be
i \mbox{Res}_{\omega=0} \sigma_{xx} = \lim_{\omega \to 0} \lim_{k \to 0} G^{xx} (\omega, k) \ .
\ee
The limit in the opposite order is related to a thermodynamic susceptibility:
\be
\lim_{k_y \to 0} G^{xx}(0,k_y) = \frac{ \partial^2 P}{\partial \xi^2} \ .
\ee
where $\xi$ is the superfluid velocity.  It follows from (\ref{Axsfexpand}) that 
\be
\frac{ \partial^2 P}{\partial \xi^2}  = \frac{ \partial J^x}{\partial \xi} = -2 \frac{  \epsilon^2}{8} \ .
\ee

As emphasized in this context in \cite{Hartnoll:2008kx}, that the limits commute implies the system really does become a superconductor below $T_c$.  Given that the limits commute, the system obeys a London type equation for $k$ and $\omega$:
\be
\langle J^x \rangle = -2\frac{  \epsilon^2}{8} A_x \ .
\ee
If we now imagine that the $U(1)$ symmetry is weakly gauged, these London equations imply not only infinite DC conductivity but also a Meissner effect with a London penetration depth that scales as $\lambda \sim 1/ \epsilon \sim (T_c - T)^{1/2}$.

\section{Heun Equation}
\label{sec:heun}

Consider a charged spin-$s$ field $h_{a_1 a_2 \cdots a_s}$ propagating in a 
curved space-time of the form (\ref{metric}) in $d+1$ dimensions,
\be
ds^2  = \frac{L^2}{u^2} \left( -f(u) dt^2 + d {\bf x}^2 + \frac{du^2}{f(u)} \right) \ ,
\ee
where ${\bf x} = (x_1 , x_2, \ldots, x_{d-1})$.
We turn on a nonzero radial electric field generated by the potential $A_t(u)$, and
an off-diagonal, purely spatial component of $h$ of the form
$\Psi(u) = h_{i_1 i_2 \cdots i_s}$ where $i$ indexes ${\bf x}$ only.  Putting aside the well known difficulties of constructing actions for charged, higher spin fields in curved space-time, 
let us suppose for the moment that this component 
$\Psi$ satisfies an equation of motion of the form
\be
\frac{1}{\sqrt{-g}} \partial_u \left( \sqrt{-g} g^{i_1 i_1} g^{i_2 i_2} \cdots g^{i_s i_s} g^{uu} \partial_u \Psi \right)
- g^{i_1 i_1} g^{i_2 i_2} \cdots g^{i_s i_s}  ( g^{tt} A_t^2 \Psi + m^2 \Psi) = 0 \ .
\ee
Given our ansatz, this equation of motion can be written in the form 
\be
\Psi''(u) + \left( \frac{f'}{f} + \frac{2s+1-d}{u} \right) \Psi'(u) + \left( \frac{A_t^2}{f^2} - \frac{m^2 L^2}{u^2 f} \right) \Psi(u) = 0 \ .
\label{basiceq}
\ee
This equation governs the onset of a superconducting instability in two different known holographic systems.  In both cases, the holographic system is in the so-called probe limit where gravity is weak and the background geometry is fixed.  In the case $s=1$ and $m=0$, the equation governs the instability in the non-abelian superconductor of  
\cite{Gubser:2008zu, GubserPufu,Roberts:2008ns}.  In the case $s=0$, the equation governs the phase transition in the s-wave superconductor of \cite{Gubser:2008px,Hartnoll:2008vx}.  
We will see in the next subsection that (\ref{basiceq}) may arise from a charged spin-2 field as well.
In the original papers \cite{Gubser:2008px,Hartnoll:2008vx,Gubser:2008zu}, the focus was on field theories in $d={2+1}$ dimensions, perhaps because of experimental relevance for high temperature superconductors which are typically layered materials.  
We will choose $A_t(u) = \mu \, (1-u^2) $ and $f(u) = 1-u^4$, implying the field theory is $d={3+1}$ dimensional.

In $d=3+1$ dimensions, the authors of \cite{Basu:2008bh} noted that (\ref{basiceq}) has an analytic solution for $\mu=4$, $s=1$, and $m=0$.  This observation was later developed by \cite{Herzog:2009ci} to calculate transport properties of the nonabelian superconductor near the phase transition.  In fact \cite{Herzog:2009ci} noted the existence of a tower of polynomial solutions for 
$\mu = 4n$, with $n$ a positive integer.

Closer scrutiny of (\ref{basiceq}) reveals that these special solutions are actually Heun polynomials
\cite{Ronveaux}.
The Heun differential equation has the form
\be
\frac{d^2 y}{dx^2} + \left( \frac{\gamma}{x} + \frac{\delta}{x-1} + \frac{\epsilon}{x-a} \right) \frac{dy}{dx}
+ \frac{\alpha \beta x - q}{x(x-1)(x-a)} y = 0 \ ,
\label{Heuneq}
\ee
where $\alpha + \beta + 1 = \gamma + \delta + \epsilon$.
The differential equation (\ref{basiceq}) can be transformed into (\ref{Heuneq}) with $a=-1$
with the identifications 
\[
\Psi(u) = \frac{u^2}{(1+u^2)^{s+k+1}} y(u) \ ,
\]
$m^2 L^2 = 4 (s-1)$, $\mu = 2 (k+s+1)$ and $x= u^2$.  One finds that
\[
\alpha = -k \ , \; \; \;
\beta = -k -s \ , \; \; \;
\gamma = s+1 \ , \; \; \;
\delta = 1 \ , \; \; \;
\epsilon = -1 -2k -2s \ ,
\]
and $q = k (k+s+1)$.
There exist polynomial solutions of this equation for all non-negative even integers $k$.  The first few polynomials $y_k$ are
\begin{eqnarray}
y_0 &=& 1 \ , \\
y_2 &=& 1 - \frac{2 (3+s)}{1+s} x + x^2 \ , \\
y_4 &=& 1 - \frac{4 (5+s)}{1+s} x + \frac{2 (54 + 25 s + 3 s^2)}{(1+s)(2+s)} x^2 - \frac{4 (5+s)}{1+s} x^3 + x^4 \ .
\end{eqnarray}

\subsection{Spin-2 fields}

The Pauli-Fierz Lagrangian for a spin two field in flat space takes the form
\be
{\mathcal L}_{\rm PF} = \frac{1}{4} \left[ h^{ab,c} h_{ab,c} - 2 h^a h_a + 2 h^a h_{,a} - {h_,}^a h_{,a} - m^2 ( h_{ab} h^{ab} - h^2 )\right] \ ,
\ee
where $h_a \equiv {h_{ab,}}^{b}$ and $h = h^a_a$.
In making this action generally covariant, ambiguities arise because of possible couplings to the Riemann curvature tensor: 
$R_{abcd} h^{ac} h^{bd}$, 
$R_{ab} h^{ac} {h^b}_{c}$, $R_{ab} h^{ab} h$, $R h^{ab} h_{ab}$,  
and $R h^2$.  
Note that while $h^c h_c$ is equivalent to $h_{ac,b} h^{b c,a}$ in flat space, in curved space, a lengthy integration by parts demonstrates that
\be
(\nabla_a h^{ac})(\nabla_b {h^{b}}_{c}) \sim
( \nabla_b h^{ac} )(\nabla_a {h^{b}}_c) + R_{ab} h^{ac} {h^{b}}_c - {R^c}_{d b a} h^{ad } {h^{b}}_{c} \ .
\ee
  
 For our spin two field coupled to a U(1) photon, we take our action to be of the form
 \begin{eqnarray}
 {\mathcal L} &=& \frac{1}{4}
 \left[
(D^c h^{ab})^*(D_c h_{ab})
- 2 ( D_a h^{ac} )^*(D_b {h^{b}}_c)
+ [ (D_b h^{b a})^* (D_a h) + c.c]
 -  (D^a h)^*(D_a h) 
\right.
\nonumber \\
&&
\hspace{5mm}
\left.
+ 2 R_{abcd} (h^{ac})^* h^{bd}
+2 R_{ab}(h^{ac})^* h{^{b}}_c
- m^2 ((h^{ab})^* h_{ab} - h^* h) 
 \right]\ ,
 \end{eqnarray}
 where $D = \nabla - i A$.  While this action is manifestly gauge invariant, it is not clear whether it is free of ghosts and tachyons.
 
We fix the only nonzero components of $h_{ab}$ to be $h_{xy} = h_{yx}$.  We take the background to have the form (\ref{metric}) and we assume that the only nonzero component of the vector field is $A_t$.  Assuming a homogenous ansatz where $h_{xy}$ depends only on the radial coordinate, $u$, the resulting equation of motion is
\be
h_{xy}'' + \left( \frac{1}{u} + \frac{f'}{f} \right) h_{xy}' + \left( \frac{ A_t^2}{f^2} - \frac{m^2 L^2}{u^2 f} \right) h_{xy} = 0 \ .
\ee
With the choice $m^2 L^2 = 4$
and $A_t = \mu(1-u^2)$, this equation will have Heun polynomial solutions for 
discrete values of $\mu$ of the form discussed above for $s=2$.\footnote{%
While this work was being prepared, ref.\ \cite{Chen:2010mk} appeared with a discussion of d-wave holographic superconductors for a different choice of action.
}

\section{Discussion}

The Abelian-Higgs model in $d=3+1$ dimensions with $m^2 L^2= -4$ we considered here was studied before numerically in ref.\ \cite{Horowitz:2008bn} as part of a survey of different models of holographic superconductivity.  We can confirm several of the numerical results in their Table 2.  The confirmation requires a little work in order to restore units and to present the results in the same statistical ensemble.  In the canonical ensemble (fixed charge density), we find
from $\mu_c = 2 \pi T$, (\ref{meanfieldgrowth}), and (\ref{Gxx}) that
\begin{eqnarray}
T_c &=& \frac{(J^t)^{1/3}}{2^{2/3} \pi}  \ , \\
\langle O_2 \rangle &\approx& - \sqrt{\frac{144}{7}} \pi^2 T_c^{3/2} (T_c - T)^{1/2} \ , \\
\lim_{\omega \to 0} G^{xx}(\omega, 0) &\approx& -\frac{72}{7} \pi^2 T_c (T_c - T) \ .
\end{eqnarray}
We have left $e=L=1$ in the above expressions and set $\A=\B=0$.
Up to a factor of two difference in the definition in $J^t$ and a minus one in the definition of $\langle O_2 \rangle$, these results agree well with 
Table 2 of \cite{Horowitz:2008bn}.

One interesting aspect of our results is their dependence on the modifications to the kinetic and potential energy of the charged scalar, parametrized by the constants $\A$ and $\B$ and the exponent $a$. 
The growth of the order parameter, the speed of second sound, and the behavior of the free energy near the phase transition are all sensitive to $A$ and $B$.  
We saw that for $K$ and $V$ of the form (\ref{KV}), if $1+4A + 4B < 0$, the second order phase transition is destroyed.  Given the existence of a second order phase transition, the critical exponents do not depend directly on the values of $A$ and $B$, but they were sensitive to the choice of
$a$ in (\ref{KValt}).  While modifications to the action of the form (\ref{KV}) seem to be generic in stringy embeddings in the large $N$ and large coupling limit, it seems likely that modifications of the form (\ref{KValt}) will be suppressed by inverse powers of $N$ and the coupling strength.\footnote{%
We would like to thank J.~Maldacena for discussion on this point.
}

The sensitivity of our results near $T_c$ on $\A$, $\B$, and $a$ suggests that as we go to lower temperature, the properties of the system will depend more and more on higher order, irrelevant terms in the gravity action.   This dependence makes our approach to superconductivity
 both richer and riskier.  Our task is richer because at the phenomenological level 
 there clearly exists a wealth of gravity theories to be explored and mined.  
 Our task is riskier because it is less clear that all of these gravity theories make physical sense interpreted as dual field theories. 
 The most well established AdS/CFT dualities are between ten dimensional string theories and supersymmetric field theories living in 3+1 space-time dimensions.  Without a stringy embedding of (\ref{Sbulk}), it is difficult to describe exactly the dual field theory or even, in fact, to be sure that such a dual exists.  
 While a generic gravity dual will possess quadratic terms in the action of the form (\ref{Sbulk}), higher order terms will depend on the details of the stringy embedding.

\section*{Acknowledgments}
I would like to thank Francesco Benini, Juan Maldacena, and Silviu Pufu for discussion.  Amos Yarom deserves special thanks both for discussion and a careful proofreading of the paper.
This work was supported in part by the US NSF under Grants No.\
PHY-0844827 and PHY-0756966.

\end{document}